\documentclass{aipproc}
\layoutstyle{6x9}

\newcommand\stitle{\footnotesize \it
\vspace{-86mm} \vspace{-2\baselineskip}
\noindent \hspace{-10mm} \mbox{
Invited talk at the "Gamma 2001" Symp. (April 4--6, 2001, Baltimore, MD).
}

\noindent \hspace{-10mm} \mbox{
AIP Conf.~Proc.~v.587, eds.~S.~Ritz, N.~Gehrels, \& C.~R.~Shrader, in press
}\vspace{80mm} \normalsize \rm}
\def\gray{$\gamma$-ray }
\def\grays{$\gamma$-rays }

\begin{document}
\title[Origin of Cosmic Rays and Diffuse Gamma Rays]{The Origin of Cosmic Rays and the Diffuse Galactic Gamma-Ray Emission}
\author{Seth W.~Digel$^1$}{address={NASA/GSFC, Code 660, Greenbelt, MD 20771
USA}}%
\footnotetext[1]{Universities Space Research Association}
\author{Stanley D.~Hunter}{address={NASA/GSFC, Code 660, Greenbelt, MD 20771
USA}}
\author{Igor V.~Moskalenko$^2$}{address={NASA/GSFC, Code 660, Greenbelt, MD
20771 USA}}%
\footnotetext[2]{NRC Senior Research Associate;~
on leave from Institute of Nuclear
Physics, M.\ V.\ Lomonosov Moscow State University, 119 899 Moscow, Russia}
\author{Jonathan~F.~Ormes}{address={NASA/GSFC, Code 660, Greenbelt, MD 20771
USA}}
\author{Martin Pohl}{address={Ruhr-Universit\"at Bochum, 44780 Bochum, Germany}}

\begin{abstract}
Cosmic-ray interactions with interstellar gas and photons produce
diffuse gamma-ray emission. In this talk we will review
the current understanding of this diffuse emission and
its relationship to the problem of the origin of cosmic rays. We will
discuss the open issues and what progress might be possible with GLAST,
which is planned for launch in 2006.
\end{abstract}

\maketitle
\stitle

\section{1. Introduction}

The problem of the origin of cosmic rays (CRs) has existed for almost 
a century.  In 1912, Victor Hess carried an electroscope aloft in 
a hot air balloon and discovered radiation increasing with altitude 
and exhibiting no day-night variation.   It had to be some kind of 
highly-penetrating radiation coming from beyond the solar system. 
Then in 1948, Freier et al. \cite{jfo:fr48} discovered that this
radiation included 
nuclei of heavy elements with relative abundances similar to the
solar system.

Much has been learned since about the important dynamical
consequences of CRs for the Galaxy \cite{jfo:pa69}.   CRs are tied to 
magnetic fields, magnetic fields are anchored in the gas phase of the 
Galaxy, and the gas is held in place by gravitational forces. 
From radio mapping of external spiral galaxies like our own Milky 
Way, CRs are known to be common to galaxies. 
Evidence for the shock 
acceleration of electrons in supernova remnants (SNR) has
recently been found (see Sec. 2).
However, no convincing evidence has been found for the commonly held 
view that CR nuclei are also accelerated in SNR.  A few 
percent of the energy released in SNR must find its way into 
energetic CRs in order to keep the Milky Way galaxy supplied with 
CRs.

Interactions of CRs with the interstellar medium (ISM) and photons
produce diffuse high-energy gamma radiation that is diagnostic of the
CRs.  The \gray fluxes are low, but \grays do have the
advantages of not being deflected by magnetic fields and low optical
depth for attenuation by intervening matter.  
Results from observations of diffuse \gray emission by the Energetic
Gamma Ray Experiment Telescope (EGRET) on the Compton Gamma Ray
Observatory have 
shown the potential of \gray measurements to contribute to solving 
the problem of the origin of CR nuclei (\cite{jfo:bea93} and references
therein).

In this paper we review the current understanding of CR 
origin and sources, CR diffusion throughout the interstellar 
medium, and the production of \grays by both CR nucleons 
and electrons.  We also review the discoveries of EGRET and explore 
the problems relating to the interpretation of those results.  We 
document some open questions and discuss how the Gamma Ray Large 
Area Telescope (GLAST)
can contribute.


\section{2. The Origin of Cosmic Rays}

Observations of the Magellanic clouds with EGRET have shown that CRs 
in the GeV range are almost certainly Galactic
\cite{mp:sr93}. However, only a few classes of objects
in the Galaxy provide sufficient energy and power to replenish the CRs,
one of which is SNR. In fact, particle acceleration
at SNR shock waves is regarded as the most probable mechanism for providing
CRs at energies below $10^{15}\,{\rm eV}$.

The distribution of SNR in the Galaxy is so poorly known and the
propagation range of
CRs, which is related to the halo size, is so weakly constrained,
that a comparison of the CR source distribution, which may be inferred
from the \gray gradient (Sec. 3), with the distribution of SNR must
be inconclusive.
A direct search for \gray emission from SNR is an alternative strategy.
In fact, a few unidentified EGRET sources are positionally coincident with
radio-bright SNR \cite{mp:es96}, but
similar correlations exist with OB associations and SNR-OB associations
(SNOBs) \cite{mp:kc96}, not to mention the possibility of these sources 
actually being
radio-quiet or highly dispersed pulsars. The
EGRET data alone do not permit a firm identification of \gray sources with SNR,
for the angular resolution is too coarse.

TeV observations using the atmospheric \v Cerenkov technique can be performed
with much higher angular resolution and sensitivity than is possible with
EGRET. However, the much higher threshold energy implies that CRs with 
energies around 100 TeV are probed. The most simple calculations
of shock acceleration \cite{mp:be78} indicate that particle spectra with 
number index $\sim2$ may be produced, but also that
acceleration cut-offs
have to be expected at 0.1--1 PeV \cite{mp:lc83}. Corresponding models of
hadronic \gray emission from SNR have suggested that a number of sources should
be detectable with present-day telescopes \cite{mp:dav94}, in particular those
embedded in dense gas \cite{mp:adv94}. Following these expectations, very deep 
surveys of TeV emission from SNR were performed, but no SNR has been
unambiguously detected as a source of
hadronic TeV \grays to date \cite{mp:bu98}. 

More careful models of shock acceleration, which include non-linear
effects arising, e.g., from the influence of the accelerated CRs
on the shocked plasma, predict particle spectra that deviate
from pure power laws \cite{mp:ba99}. Also, the energy and momentum carried
by electromagnetic turbulence and the motion of the CR scattering 
centers relative to the plasma modify the process of shock acceleration,
such that an $E^{-2}$ spectrum is not the canonical result it
was thought to be \cite{mp:lps00}. In fact a distribution of spectral indices should exist, and this is actually observed in the radio
spectra of shell-type SNR \cite{mp:gr00}. Therefore, the non-detection
of hadronic TeV
emission from SNR may be not incompatible with CR acceleration in
these sources;
nevertheless, the results seem to conflict with the notion that SNR
accelerate CR hadrons up to the ``knee'' (Sec. 3).

The recent detections of non-thermal X-ray synchrotron radiation
from the SNR SN1006 \cite{mp:koy95}, RX 
J1713.7-3946 \cite{mp:koy97,mp:sl99}, IC443 \cite{mp:keo97}, Cas A \cite{mp:al97}, and RCW86 \cite{mp:bo01},
and the subsequent detections of SN1006 \cite{mp:ta98}
and RX J1713.7-3946 \cite{mp:mur00} at TeV energies 
support the hypothesis that at least Galactic CR electrons are 
accelerated predominantly in SNR.
The relative intensities of the keV synchrotron emission and the TeV
inverse Compton (ICS) radiation are independent of the acceleration process and
the model thereof \cite{mp:po96}. Care has to be
exercised in separating thermal from non-thermal X-ray emission \cite{mp:dr01},
though, 
and the substructure of the SNR shock as well as propagation effects have
to be taken into account when interpreting the TeV data \cite{mp:aa99,mp:at00}.    Nevertheless SN1006 and RX J1713.7-3946 are
obvious examples of leptonic TeV \gray emission. The most recently 
detected SNR, Cas A \cite{mp:aha01}, is a less clear case, but presumably
the emission is also of leptonic origin \cite{mp:ato00}.

It is interesting to note that for all SNR the X-ray flux, synchrotron or not, 
is less than the extrapolated radio synchrotron spectrum. Since many
of the sources, in particular the five historical remnants, are too young for
their electron spectra to be limited by energy losses, acceleration cut-offs 
must occur at electron energies of 100 TeV or less \cite{mp:rk99}, which would be intrinsic to the actual acceleration process.

The production of CR electrons in SNR has important
consequences. At energies higher than about 100 GeV the 
lifetime of electrons, and thus
their range, is rather short; only a few SNR would contribute to the 
locally observable CR electron spectrum. Therefore the locally
measured spectrum would not be
representative of the spectrum elsewhere in the Galaxy.
This conclusion is the basis for the ICS models of the GeV excess
\cite{mp:pe98} (Sec. 3).

\section{3. Propagation of Cosmic Rays}

The spectrum of CRs can be approximately described by a single power law with
index --3 from 10 GeV to the highest energies ever detected,
$\sim10^{20}$ eV. The only feature observed is a ``knee'' around
$10^{15}$ eV. Because of this featureless spectrum, CR production and
propagation are believed to be governed by the same mechanism
over decades of energy; a single mechanism works below the knee
and the same or another one works above the knee, although the origin
of the CR spectrum is not still understood.

Energetic CR interactions
with gas or magnetic and radiation fields in the interstellar medium
produce \grays that carry
information about these interactions such as the
spectrum and flux of CR and physical conditions such as the
gas density and radiation field.
Some portion of \grays is produced near the
CR sources, like SNR and pulsars, that generally appear as point sources
to \gray telescopes.
The rest are produced in CR interactions in the ISM and are therefore
diffuse.

In the ISM particles
diffuse and lose or gain energy, and so their spectra change from
their initial forms.
Freshly-accelerated particles (Sec.\ 2) propagate in the ISM where
they produce secondary particles and $\gamma$-rays.
Electrons produce synchrotron photons and \grays via ICS
and bremsstrahlung.
Nucleons spallate and produce secondary nucleons,
antiprotons, and charged pions that give rise to secondary positrons
and electrons.  Decays of secondary $\pi^0$'s also produce
$\gamma$-rays.

CRs can be measured directly only in the solar system,
in the outskirts of our Galaxy.
Those observed in the solar system are a complicated mixture
of primary and secondary particles which are diffusing through
the ISM from their sources.  Only photons or \grays are
able to deliver the information directly from other parts of the
Milky Way, but this information is
integrated over the line of sight.
To extract information about CRs from \grays models of CR propagation
are needed.

CRs propagate throughout the Galaxy guided by magnetic fields.
They are held in the Galaxy by scattering on magnetic irregularities,
the process described mathematically as diffusive propagation.  The models
can be complex and distinguish between the diffusion rates in the thin
galactic disk and extensive halo \cite{im:jones01}.
The diffusion coefficient may
depend on the local plasma conditions and is probably different in the
disk and the halo.  It is often assumed to depend on particle rigidity
with index 0.33--0.6, with the former value reflecting a ``Kolmogorov''
spectrum of magnetic scattering centers and the latter obtained
phenomenologically from fitting the measured B/C ratio.  Propagation
may be influenced by a galactic wind
(convection of particles away from the Galaxy) and/or diffusive
re-acceleration (a second order Fermi process).

The diffusive model can be shown to
be equivalent to, and because of the complexities mentioned above is
often replaced by, a simpler empirical formalism
known as the ``leaky-box model.''  
In this model the principal parameter is an effective escape length or
grammage (column density of matter traversed, in g cm$^{-2}$) and the
sources and particles are uniformly distributed in space and time
throughout the Galaxy.  The grammage parameter is determined by fitting
to the data and found to be a broken power law in rigidity,
increasing at low energies and decreasing with index --0.6 at
relativistic energies.



Detailed CR diffusion models have been developed over the last
30 years. The GALPROP model \cite{im:SM98}, for example, is a
recent numerical model
which combines the Galactic structure with diffusive reacceleration and
convection in the ISM. This model incorporates all CR species with $Z<29$
including leptons and antiprotons, together with \grays and
synchrotron emission.

Measurements of primary CR abundances or stable secondary/primary ratios
do not permit distinguishing between the leaky box and diffusion models.
However, the propagation of radioactive secondary isotopes depends on the
lengths of time that particles spend in the disk (production) and the halo
(decay), therefore making them sensitive probes of the propagation model.

 
\begin{figure}[tbp]
\includegraphics[width=.494\textwidth]{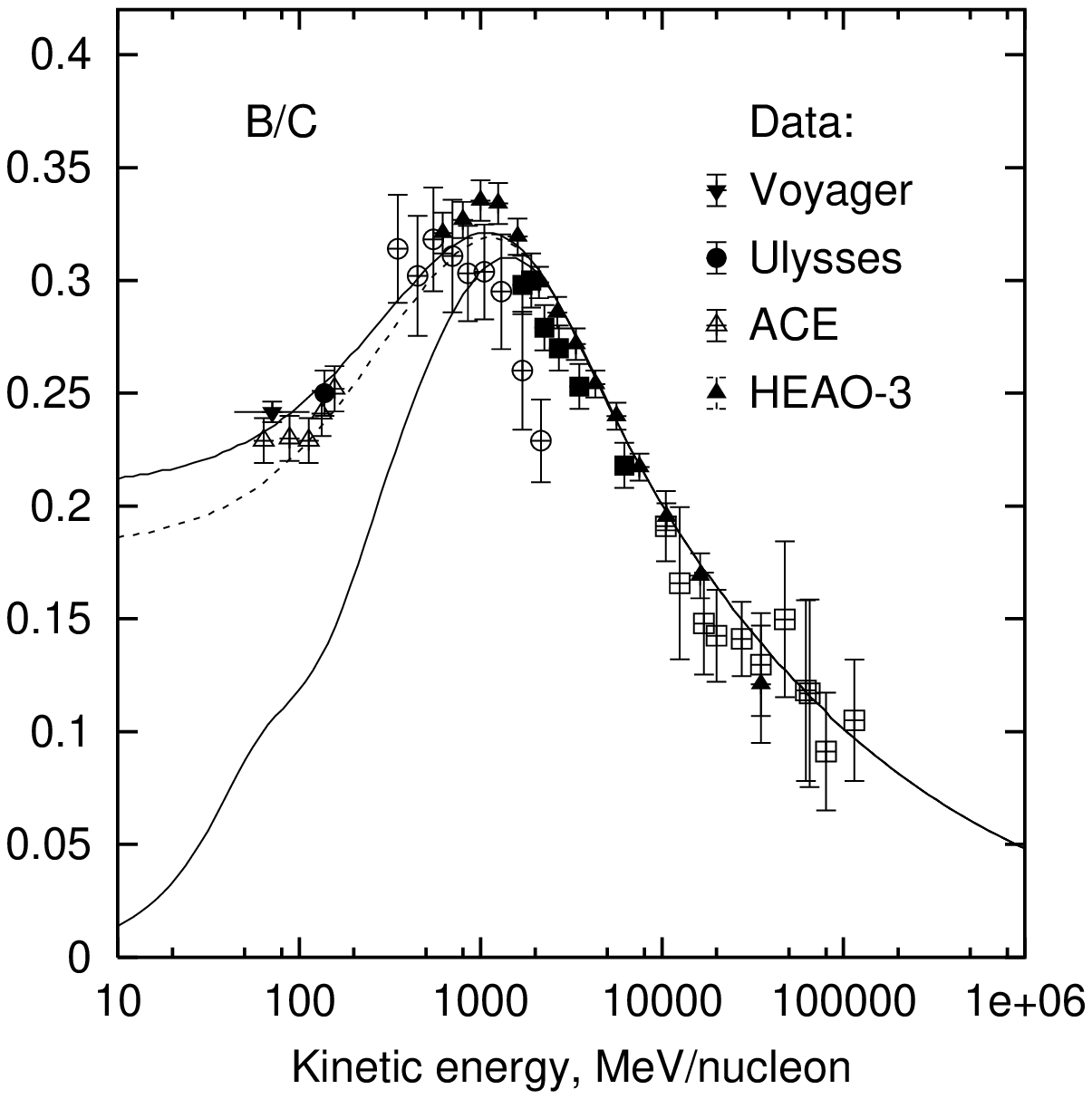}
\includegraphics[width=.55\textwidth]{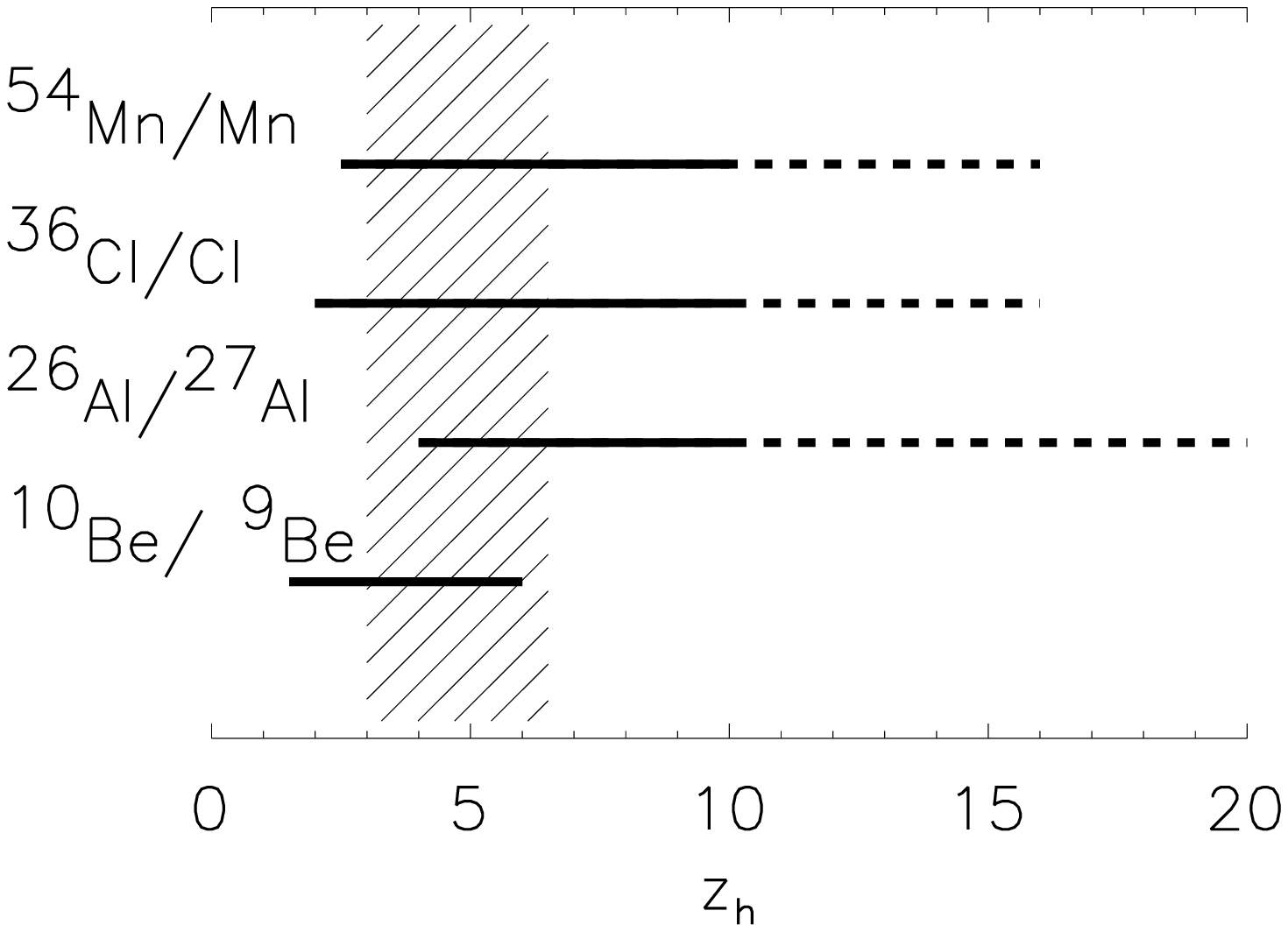}
\caption{\underline{\bf Left:}
B/C ratio calculated for CR halo scale height $z_h=4$ kpc \cite{im:SM01}.
Lower curve local interstellar spectrum, upper modulated: solid curve --
$\Phi=500$ MV,
dotted curve -- $400$ MV.
Data below 200 MeV/nucleon: ACE \cite{im:davis},
Ulysses \cite{im:duvernois}, Voyager \cite{im:lukasiak};
high energy data: HEAO-3 \cite{im:engelmann},
for other references see \cite{im:SS}.
\underline{\bf Right:}
Halo size limits \cite{im:SM01} derived from the abundances of the
4 radioactive isotopes and ACE data.
The ranges reflect errors in ratio measurements, source abundances,
and production cross sections. Dashed lines indicate uncertain upper limits.
The shaded area indicates the range consistent with all ratios.}
\end{figure}

Propagation parameters are usually derived using B/C and (Sc+Ti+V)/Fe
ratios, while radioactive secondary isotopes $^{10}$Be, $^{26}$Al, $^{36}$Cl,
$^{54}$Mn (all with $T_{1/2}\sim$ 0.3--2 Myr) allow the size of the halo to
be constrained.
Figure 1 shows examples of B/C calculations and halo size constraints
derived from radioactive isotopes \cite{im:SM01}.
Once defined from B/C and Be measurements, the propagation
parameters determine other isotopic ratios.

Typical values of the diffusion coefficients depend on the propagation
model and values of cross sections employed, but all of them are of the order
of few times $10^{28}$ cm$^2$ s$^{-1}$. Larger values make the
propagation more rapid and produce smaller amounts of secondaries, smaller
values increase the secondary/primary ratios.



The spectrum of diffuse \grays from the inner Galaxy has excesses
at high and low energies compared to calculations based on the
assumption that the nucleon and electron spectra do not change shape
throughout the Galaxy \cite{im:hunter,im:SMR00} (see Sec. 4 and Fig. 3).
This implies that the local spectra may be not representative
for the Galaxy as a whole. At low energies, the excess may be due to
unresolved point sources that dominate in the MeV range and below
\cite{im:azita}.
At high energies, where most photons come from ICS and $\pi^0$ decay,
the excess can be explained by spectra of
nucleonic \cite{im:mori} and/or leptonic \cite{mp:pe98} components that
are harder than those observed locally.

Secondary antiprotons and positrons are produced in the same nucleonic
interactions in the ISM as $\pi^0$'s and thus provide information
complimentary to that of \grays (without direction information), but
in this case the agreement is good
\cite{im:SMR00}. The leptonic hypothesis of the origin of the excess at
high energies therefore may be more plausible,
especially because of large energy losses of high energy electrons.


A new test for the origin of the excess will be feasible with
the new generation telescope GLAST (Sec. 5), which will be able to
measure \grays up to 300 GeV for the first time.
While improved angular resolution will allow better discrimination
of the
point source contribution, the capability for spectral measurements
at sub-TeV energies is essential to distinguish between diffuse
nucleonic and ICS spectral components.
The shape of the \gray spectrum from $\pi^0$-decay resembles
the spectrum of the nucleonic component of CRs; the ICS spectrum is
flatter and its cut off energy is determined by the maximum energy
to which SNR can accelerate electrons. Because this energy is in the
1--100 TeV range, the spectrum of \grays in the 10--100 GeV range
is therefore crucial \cite{im:SM}.


\section{4. Results and Questions from EGRET}

\begin{figure}[tbp]
\resizebox{0.78\textwidth}{!}
  {\includegraphics{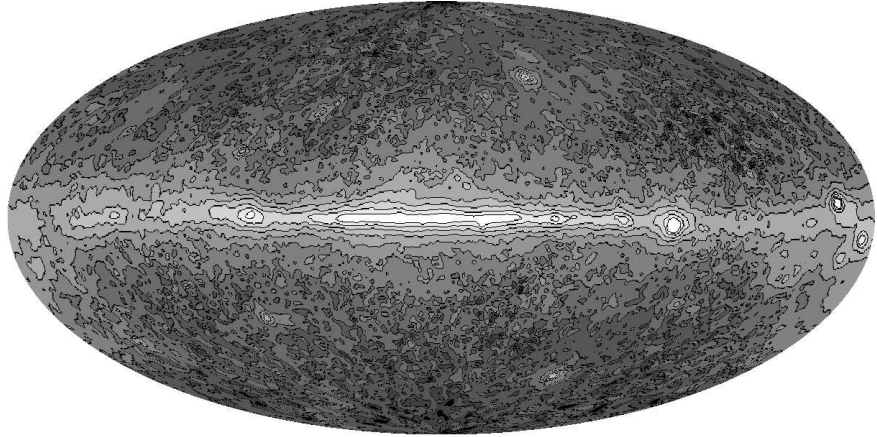}}
\caption{Intensity of gamma rays ($>100$ MeV) observed by EGRET.
The broad, intense band near the equator is interstellar
emission from the Milky Way. The intensity scale ranges from $1 \times
10^{-5}$ cm$^{-2}$ s$^{-1}$ sr$^{-1}$ to $5 \times 10^{-4}$ cm$^{-2}$ s$^{-1}$
sr$^{-1}$ in ten logarithmic steps.  The data have been smoothed slightly
by convolution with a gaussian of FWHM 1.5$^\circ$.}
\end{figure}

EGRET completed the first all-sky survey at high-energies ($> 30$ MeV,
Fig. 2).
More than 60\% of the \grays that EGRET detected are from CR interactions in
the Milky Way.  The sensitivity as well as the excellent background rejection of
EGRET enabled great progress in the study of interstellar \gray emission.

EGRET observations of the Large and Small Magellanic clouds confirmed
that CRs are galactic in origin, rather than metagalactic or universal. 
The \gray flux of the LMC measured by EGRET is consistent with a CR density 
similar to that in the Milky Way \cite{hu:sr92}. However,
the upper limit measured by EGRET for the SMC implies a CR density several times
less \cite{mp:sr93}.

The EGRET \gray spectrum of the inner Milky Way (Fig. 3) provided the first clear
evidence for both electron and proton CRs across the Galaxy, with the spectrum
following the $\pi^0$ ``shoulder'' for energies greater than the proton emissivity peak
at half the $\pi^0$ rest mass \cite{im:hunter}.
Above $\sim100$ MeV, $\pi^0$ decay \grays from proton-nucleon interactions are the 
dominant spectral component.  At lower energies the spectrum is dominated by electron 
interactions via bremsstrahlung and ICS.

Models of the interstellar gamma-ray emission of the Milky Way based on the
known \gray production mechanisms, together with inferred distributions of
interstellar gas,
low-energy photons, and CRs, predict \gray intensities consistent with the
observations on scales of degrees \cite{im:SM96,im:hunter,mp:pe98,im:SMR00}.
Owing to the
limited statistics and angular resolution of the data, such models are essential
for determining accurate positions and fluxes of \gray point sources at low
latitudes.  They are also the means to discover the distribution of CRs in
the Milky Way.  Broadly speaking, the CR distributions derived from the models are
consistent with each other.  However, differences in approach, including
techniques used to
resolve the non-unique inversion of radio and millimeter spectral line surveys
into the 3-dimensional distribution of gas, presently limit the usefulness of
detailed comparisons between models.

\begin{figure}[tbp]
\resizebox{0.895\textwidth}{!}
  {\includegraphics{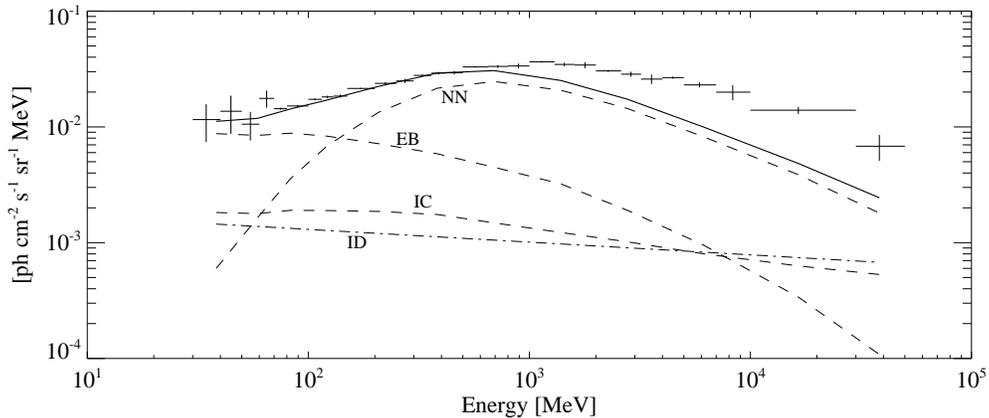}}
\caption{Spectrum of the inner Milky Way ($|l| < 60^\circ$, $|b| < 10^\circ$)
with calculated components from bremsstrahlung (EB), inverse Compton (IC),
$\pi^0$ decay (NN), and extragalactic isotropic emission (ID)
\cite{im:hunter}.
}
\end{figure}


For the nearest interstellar clouds, EGRET provided linear resolutions of $\sim
10$ pc.  Studies of these, e.g.  \cite{hu:di99}, permit the
tightest direct constraints on confusion of unresolved
gamma-ray point sources with diffuse emission, and also on the
assumptions that CR densities and
the molecular mass calibration, i.e., the relation of 2.6 mm CO line
intensity $W_{\rm CO}$ to $N({\rm H}_2)$, are uniform on the scale
of interstellar clouds.

The GeV excess (see Sec. 3) is evident in Fig. 3, as well as in spectra
for smaller angular scales. 
Several explanations of the excess emission are possible:  EGRET
calibration error,
uncertainty in the $\pi^0$ production, unresolved hard point sources,
and a Galactic average proton and/or electron spectrum that may be harder
than observed locally.
An explanation involving ICS from electrons is perhaps the most plausible (see Sec. 3).
Calibration error is unlikely, as the observed spectra of EGRET point sources,
which are generally 
well described by single power laws, do not harden above 1 GeV
(e.g., \cite{hu:ul95}).  The $\pi^0$ production function has recently been
re-evaluated using nuclear interaction Monte Carlo codes,
and the hardening of emissivity predicted at GeV energies is not 
sufficient to explain the GeV excess \cite{hu:ch01}.
Hunter et al. \cite{im:hunter} concluded from the shape of the spectrum of the
inner Galaxy that the contribution from unresolved sources with
power-law spectra
is $< 10$\%.  This estimate is, however, rather uncertain
because a large contribution from unresolved sources distributed closely
like the
molecular gas -- although not evident in studies of local clouds -- could
be accounted for by reducing the $N({\rm H}_2)/W_{\rm CO}$ ratio.
Pulsars not detected individually could contribute significantly above 1 GeV,
although apparently not with the correct latitude
distribution \cite{sd:pea97, hu:zh98}.

A halo of high-energy ($>100$ MeV) \grays about the Milky Way remains
when the EGRET team's
interstellar emission model is subtracted from the observations
\cite{hu:di98,hu:sr98}.  The simplest interpretation for the halo is
incomplete accounting of the ICS emission at high latitudes
(e.g., \cite{hu:mo00}).
Other interpretations based on CR interactions in the halo
with very cold H$_2$ in dense clumps, a dark matter candidate,
have been proposed (e.g., \cite{hu:de99}).  The ICS interpretation
alone appears sufficient.  EGRET data may in fact offer little insights about
baryonic dark matter in the
halo.  Even if the dark matter is in dense clumps, they
could be dense enough to attenuate any \grays produced in them \cite{hu:ka99}.

\section{5. Advances Anticipated with GLAST}

The general capabilities of the Large Area Telescope on the Gamma-ray
Large Area Space Telescope (GLAST), which is planned for launch in 2006,
are described elsewhere in this volume \cite{sd:ge01}.  The advantages of the large
effective area, high observing efficiency, and narrow point-spread function
of GLAST for the study of CR production and diffuse \gray emission
are considered here.

Supernova remnants have been established as acceleration sites of
leptons, but the acceleration of hadrons has not yet been
detected (Sec. 2).  The observational signature of proton acceleration would
be expected to be $\pi^0$-decay \gray emission from proton collisions
with nucleons in interstellar clouds being overtaken by SNR shocks.
The \gray spectrum would appropriately reflect the hard spectrum of
the CRs.

Although several EGRET \gray sources are spatially
coincident with SNR \cite{mp:es96,sd:sd95}, the
large positional uncertainties for the EGRET sources and the
limited photon statistics do not permit $\pi^0$-decay
emission from the shell to be distinguished from leptonic emission
processes, or even from \gray emission from associated pulsars.

\begin{figure}[tbp]
  \includegraphics[width=1.0\textwidth]{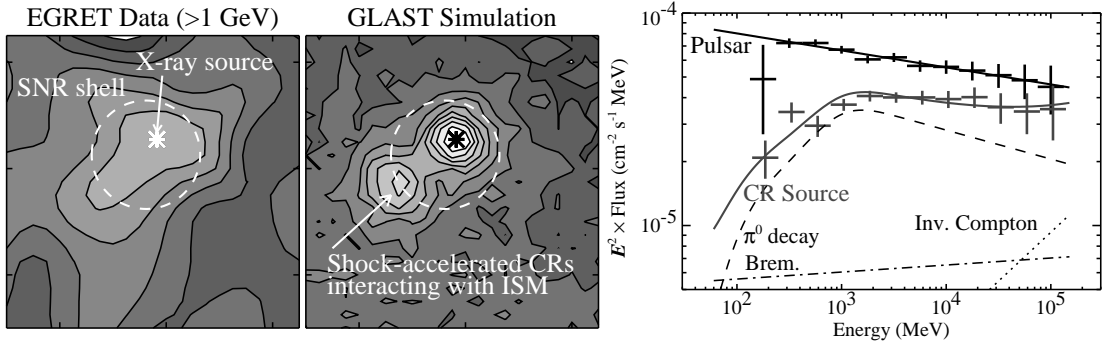}
\caption{EGRET observation (summed Phase 1--4) and GLAST simulation
(1-year sky survey) of the $\gamma$-Cygni SNR.  The dashed circle is the
location of the shell \cite{sd:hi77}.  The spacing of the tick marks is
$1^\circ$.  Also shown are simulated measurements with GLAST of the
spectra of the pulsar and CR source.  See text for discussion of the
$\gamma$-Cygni model.}
\end{figure}

The potential for GLAST to clarify the nature of the associations is
illustrated in Figure~4, which shows EGRET data and a GLAST simulation
for the $\gamma$-Cygni SNR.  This region was detected as a point source by
EGRET (and designated as 3EG J2020+4017 \cite{sd:hea99}).  An X-ray
source at the position indicated in the figure has the spectral
characteristics of a pulsar and has been proposed as the counterpart
to the EGRET source \cite{sd:bea96}.  For the simulation, it
was assumed that the flux could be divided 60\%/40\% between the
prospective pulsar and a source at the shell of the SNR where the
EGRET data suggest an extension.  Spectra for the two sources were
selected to be consistent with the overall spectrum of 3EG~J2020+4017
and the spectral components of the shell source were chosen to be
consistent with models of the \gray emissivity of SNR
\cite{sd:gps98} (for details see \cite{sd:ado99}).  The GLAST
intensity map shows that the CR source will be resolved from the pulsar
at high energies.
In addition, the spectra of the sources will be separately measurable
at energies above 150 MeV, and the $\pi^0$-decay component of the
shell source will be strongly detected.

For the study CRs in the Milky
Way, the superior effective area and angular resolution of GLAST will
permit important advances.  GLAST will reveal whether unresolved (by EGRET)
point sources are
only a minor component of the celestial \gray flux, as is
currently suspected (Sec. 4; \cite{im:hunter}).
The angular resolution of GLAST also will be important for studying the
coupling of CRs to the spiral arms in the inner Milky Way.
The diffuse interstellar emission is intense and quite structured near
the Galactic equator, and evidence for coupling will require careful
separation of components at different distances, most likely by study
of the emission near the tangent directions of the arms.

GLAST observations of interstellar clouds at GeV energies may permit
a new test of the origin
of the GeV excess.  CR electrons fully penetrate molecular clouds and
produce \grays via bremsstrahlung and ICS with low-energy 
(microwave to ultraviolet) photons within the cloud.  The radiation
field inside a cloud varies with visual extinction to its center:
the ultraviolet and optical radiation intensities decrease rapidly with increasing depth
whereas the far-infrared radiation is $\sim 10$ times
more intense throughout the cloud 
than the Galactic interstellar radiation field \cite{hu:ma83}.  The resulting
variation of the gamma-ray spectrum across the face of a cloud could be used
to determine the ICS component.

GLAST will map the interstellar \gray emission of the LMC, SMC, and
possibly M31, and additionally will likely detect several other local
group galaxies and starburst galaxies as point sources, greatly expanding the
potential for high-energy \gray studies of CRs.

\section*{Acknowledgements}

IVM acknowledges support from the NRC/NAS Research Associateship Program.
MP acknowledges partial support by the Bundesministerium f\"ur Bildung und
Forschung, grant DLR 50 QV 0002.

\end{document}